\documentclass[useAMS,usenatbib]{mn2e}
\usepackage{graphicx}
\usepackage{lineno}	

\title[Periodicities in the coronal rotation and sunspot numbers] {Periodicities in the coronal rotation and sunspot numbers}

\author[Satish Chandra and Hari Om Vats]{Satish Chandra$^{1}$\thanks{
satish0402@gmail.com} and Hari Om Vats$^{2}$\thanks{vats@prl.res.in}\\
$^{1}$Department of Physics, PPN College, Kanpur - 208 001, INDIA.\\
$^{2}$Physical Research Laboratory, Ahmedabad - 380 009, INDIA.}
\begin{document}

\date{Accepted 8888 XXX 88 Received 8888 XXX 88; in original form 8888 XXX 88}

\pagerange{\pageref{firstpage}--\pageref{lastpage}} \pubyear{8888}

\maketitle


\label{firstpage}

\begin{abstract}
The present study is an attempt to investigate the long term variations in coronal rotation by analyzing the time series of the solar radio emission data at 2.8 GHz frequency for the period 1947 - 2009. Here, daily adjusted radio flux (known as \emph{Penticton flux}) data are used.  The autocorrelation analysis shows that the rotation period varies between 19.0 to 29.5 sidereal days (mean sidereal rotation period is 24.3 days). This variation in the coronal rotation period shows evidence of two components in the variation; (1) 22-years component which may be related to the solar magnetic field reversal cycle or Hale's cycle, and (3) a component which is irregular in nature, but dominates over the other components. The crosscorrelation analysis between the annual average sunspots number and the coronal rotation period also shows evidence of its correlation with the 22-years Hale's cycle. The 22-years component is found to be almost in phase with the corresponding periodicities in the variation of the sunspots number. 
\end{abstract}


\section{Introduction}

Coronal rotation can be observed through various solar tracers at different frequencies, like coronal green line (Fe XIV emission line at 530.3 nm ), white light, He I line (at 1083 nm), soft X-rays, UV rays, radio waves. The coronal green line has been used to measure the rotation rate of the solar corona at higher latitudes by \citet{Waldemier50, Trellis57, Cooper62, Sykora71, Sime89, Rybak94, Badalyan06a, Badalyan06b} and others. The results of \citet{Waldemier50} and \citet{Cooper62} indicate a faster rate of rotation as compared to the rate of rotation of the sunspots, suggesting a much lower differential rotation rate in the corona. In his work on green corona \citet{Sykora71} found that the Sun shows little or no differential rotation for six latitudinal zones $\pm 7.5$, $\pm 27.5$ and $\pm 47.5$. For low latitudes the rotation period was near to that found by \citet{Trellis57}. The green (Fe XIV at 530.3 nm) emission line for the period 1973 - 2000 and red (Fe X at 637.4 nm) emission line for the period 1984 - 2000 were  analyzed by \citet{Altrock97, Altrock03}. It was reported that the corona, at green and red emission lines, shows more rigid rotation than does the photosphere. \citet{Sime89} also concluded, after analyzing the Sacramento Peak Observatory (SPO) data observed between 1973 and 1985, that the Fe XIV corona rotates more rigidly than do features in the photosphere or chromosphere. The synodic period obtained by \citet{Rybak94} for the period 1964 - 89 again confirmed the differential rotation of the green corona. \citet{Badalyan06a, Badalyan06b} carried out a comprehensive analysis using a long database (1939 - 2001) on the brightness of the coronal green line. The results support previous conclusions that the differential rotation in the corona is less pronounced than in photospheric tracers.

\begin{figure*}
  \centering{
  \includegraphics[width=\textwidth]{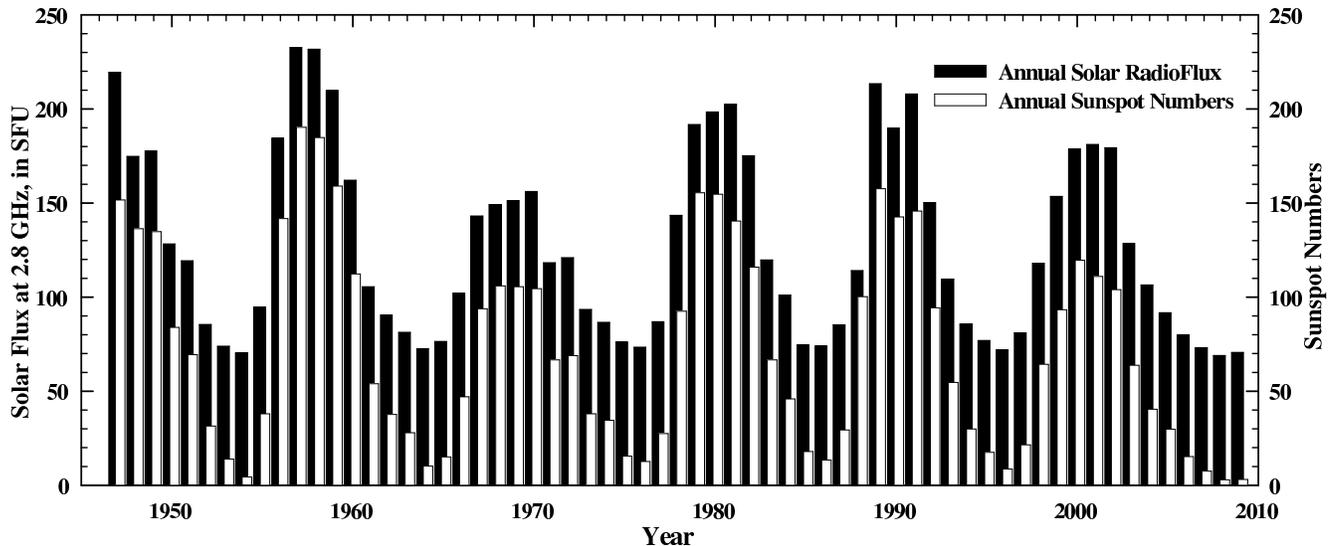}}
 \caption{The annual mean of solar radio flux at 2.8 GHz is compared with the annual mean of sunspot numbers for the period 1947-2009.}
 \label{srf1}
 \end{figure*}

\citet{Hansen69} used the K-coronometer for coronal rotation measurement at different latitudes, for heights ranging from 1.125 to 2 $R_{\sun}$. The rotation found at the equator is in good agreement with the sunspot's rotation results and shows less variation with latitude at higher latitudes in comparison to the rotation of the chromosphere. A detailed study of the white light corona, from 1.1 $R_{\sun}$ to 30 $R_{\sun}$, was done with the \emph{LASCO} coronagraphs on board the \emph{SoHO} spacecraft. It was concluded that the rotation of the corona displayed a radially rigid rotation of 27.5 days synodic period from 2.5 $R_{\sun}$ to $>$ 15 $R_{\sun}$ \citep{Lewis99}.

The He I 1083 nm maps, from the National Solar Observatory (NSO), have been used to determine the rotation. It is found, both from observations and from magnetic extrapolation methods, that the corona becomes more rigid with height. By considering coronal holes as tracers (from an atlas of coronal holes mapped in He I 1083 nm data) of the differential rotation \citet{Insley95} demonstrated that the mid-latitude corona rotates more rigidly than the photosphere, but still exhibits significant differential rotation, with an equatorial rate of 13.30 $\pm$ 0.04 deg/day, and at 45\degr latitude a rate of 12.57 $\pm$ 0.13 deg/day. An analysis of the rotation of coronal holes spanning 18 years (from 1973 to 1991) was done based on data from the Catalogue of Coronal Holes \citep{Navarro94}. Isolated coronal holes showed a typical differential rotation, but polar coronal holes extensions displayed two different types of behavior: a rotation rate below approximately $40\degr \pm 5\degr$ of heliographic latitude, increasing to the equator, and a rotation rate above the same heliographic latitude but increasing to the poles.

Coronal holes, as observed from the \emph{Skylab} and \emph{Yohkoh} spacecrafts, have also been used to determine the rotation rate of the outer corona. Soft X-ray (SXR) observations of an elongated coronal hole, shows the almost rigid rotation of the coronal hole \citep{Timothy75, Kozuka94}. The solar full disc (SFD) images, obtained by soft X-ray telescope (SXT) of \emph{Yohkoh} space observatory, were used by different scientific groups to study the rotation rate of the corona. \citet{Weber99, Weber02} and \citet{Chandra10} concluded, after analyzing SXT data by different methods, that the rotation profile of the corona across the latitude is shallower than the rotation profile of its lower atmospheric levels. \citet{Kariyappa08} tracked the X-ray bright points (XBPs) on SFD images observed through the SXT and XRT (X-ray telescope) on board the \emph{Yohkoh} and \emph{Hinode} spacecrafts, respectively. \citet {Kariyappa08} found, contrary to all expectations, that the corona rotates differentially with respect to latitude, as in the case of the photosphere and the chromosphere.

\citet{Karachik06} analyzed the coronal bright points (CBPs) on SFD filtergrams observed through \emph {SoHO}/EIT (Fe XII line at 19.5 nm ) and reported that the rotation of CBPs closely follows the latitudinal rotation profile of the photospheric magnetic field. It was also shown that coronal features at different heights in the corona exhibits different rotation rates. \citet{Brajsa02, Brajsa04, Mulec07} and \citet{Brajsa08} determined the solar differential rotation by tracing coronal bright points on SFD filtergrams observed through \emph {SoHO}/EIT (Fe XV line at 28.4 nm ). For the declining phase of solar cycle 23, \citet{Zaatri09} compared differential rotation of subphotospheric layers derived from GONG++ dopplergrams with the small bright coronal structures (SBCS) observed through \emph {SoHO}/EIT . It is found at the equator that the SBCS rotate faster than the upper subphotospheric layer (3Mm) by about 0.5 deg/day. The latitude gradient of the rotation rate of the SBCS and the subphotospheric layers were found in agreement.

\citet{Kane01, Vats01} used the disk integrated solar radio fluxes for the period 1997 to 1999 at different frequencies to study the coronal rotation at different heights. \citet{Vats01} shown that coronal rotation depends on the heights in the corona. The autocorrelation analysis of radio SFD images at 17 GHz, for the period 1999 - 2001, also ascertained that the solar corona rotates less differentially than does the photosphere and the chromosphere \citep{Chandra09}.

\citet{Rybak94} reported that the differential rotation of the green corona shows no clear cyclic variation of the rotation. \citet {Kariyappa08}, who tracks the XBPs on the SFD images of \emph{Yohkoh}/SXT and \emph{Hinode}/XRT, also found that the rate of coronal rotation is independent of the phases of the solar activity cycle. But, from the GONG (Global Oscillation Network Group) and MDI (Michelson Doppler Imager) data, \citet{Jain01}, have shown that the equatorial rotation rates show variation with the solar activity cycle. \citet{Badalyan06a, Badalyan06b} also reported that there is periodic variation in the rate of differential rotation of green corona which is related to the solar activity cycle, a correlation usually found in photospheric tracers \citep{Hathaway90}. With the SFD filtergrams obtained from \emph{SoHO}/EIT (Fe XV at 28.4 nm), \citet{Brajsa04} exhibited that the differential rotation profile corresponds roughly to the rotation of sunspot groups. \citet{Mulec07} and \citet{Brajsa08} found, with the same data of different epoch, that the equatorial rotation velocity and the gradient of differential rotation show similar values during periods of low and high activity. 

\citet{Mouradian02} analyzed the 2.8 GHz radio emission flux with the maximum entropy method and showed that the 2.8 GHz radio emission rotation varies according to the activity level. But \citet{Mehta05}, after analyzing the radio emission data could not find any systematic relationship between the coronal rotation period and the phase of the solar cycle. \citet{Chandra09, Chandra10} calculated the equatorial rotation rate for each year separately (using radio images at 17 GHz for the years 1999-2001 and SXT data for the years 1992-2001, respectively) and compared it with the annual sunspot numbers. The comparison shows that the equatorial rotation rate seems largely to be a function of the phases of the solar cycle.

\citet{Javaraiah00} used the GPR (Greenwich Photoheliographic Results) data on sunspot groups compiled during 1879 - 1975 to study the variations of the differential rotation coefficients $A$ and $B$ during the odd numbered solar cycles (ONSCs) and during the even numbered solar cycles (ENSCs). The parameters $A$ and $B$ are measures of the equatorial rotation rate and latitude gradient of rotation rate, respectively. It is seen that the variation in $A$ is significant only in the ONSCs (Waldmeier cycle number 13, 15, 17 and 19). Whereas, the variation in $B$ is quite significant in both ONSCs and ENSCs (12, 14, 16, 18 and 20). There exists a good anticorrelation between the mean variation of $B$ during ONSCs and ENSCs, suggesting the existence of a 22-years periodicity in $B$.

Different methods and different tracers have been employed to determine the coronal rotation and its dependence on the phases of the solar cycle, but the estimates were not found to be in agreement with each other. Using radio measurements to estimate solar rotation is not very new, but since the last couple of decades many scientific groups have employed radio flux data at various frequencies to ascertain the rotation period \citep{Vats98a, Kane01, Mouradian02}, its differentiality as a function of altitude \citep{Vats01} and latitude \citep{Chandra09} and its relationship with the phases of the solar cycle \citep{Mehta05}. The present study is an attempt to estimate the coronal rotation rate and its relationship with the solar magnetic cycle by analyzing the time series of the radio emission data at 2.8 GHz frequency (wavelength 10.7 cm) for the period 1947 - 2009. The data covers almost six solar activity cycles and is currently one of the best and longest data of solar indices we have after the sunspot number. This radio emission at 2.8 GHz frequency originates deep in the solar corona and hence represents coronal rotation \citep{Vats01}. 

The next section discusses the observation and importance of radio flux data at 2.8 GHz used in the present analysis. Its subsection briefly describes the autocorrelation and crosscorrelation analysis technique, primarily used to determine the rotation period and its correlation with the solar activity, respectively. The paper ends with a discussion of the results obtained.

\section{Data}

The data used in the present work comprises disc integrated flux measurements of radio waves at 2.8 GHz. It is expressed in solar flux units ($1$ SFU =$10^{-22}$ Watts/meter square-Hertz). It is almost thermal in origin because emission is mainly due to thermal free-free (bremsstrahlung) radiation \citep{Tapping90}. It is directly related to the total amount of magnetic flux. The data measurements have been continued through the Solar Radio Monitoring Programme, a service which is operated by the National Research Council of Canada. The solar flux density at 2.8 GHz is measured using two radio telescopes (called flux monitors). The two instruments record the strength of the solar radio emission at 2.8 GHz frequency each day for as long as the Sun is above the horizon. Between 1946 and 1990, the measurements were made in the Ottawa area at the Algonquin Radio Observatory. In 1990, following the closure of that observatory, the system was relocated to the Dominion Radio Astrophysical Observatory, near Penticton, British Columbia. It now forms a consistent, uninterrupted data base covering more than 63 years. The observations, which started in late 1946, are still continuing. The continuity and consistency of calibration of these measurements of the 2.8 GHz radio flux have helped to establish it as an internationally accepted index of solar activity.

These data are readily available through the National Geophysical Data Center (NGDC), NOAA, USA. The data
available at the site of NGDC are tabulated in two forms; the observed flux and the adjusted flux. The observed flux data are the actual measured values and are affected by the changing distance between the Earth and Sun, whereas the adjusted flux data are scaled to the standard distance of 1 AU. Therefore, the adjusted flux data are considered to be more appropriate for the study of the behavior of the solar corona. The annual mean of the daily measurements of the radio emission can be represented in the form of a time series (shown by a solid line in Figure \ref{srf1}). This figure shows the yearly variation of sunspots number as well as the radio emission at 2.8 GHz. 

\citet{Covington69} made a comparison over more than a solar activity cycle and showed that there is indeed a linear correlation between the 2.8 GHz solar flux and the total photospheric magnetic flux in the active region. Various manifestations of solar activity are driven by the total amount of magnetic flux emerging through the photosphere into the chromosphere and the corona, and its spatial and temporal distribution. Besides this, the 2.8 GHz solar flux correlates well with some other indices of solar activity such as sunspot number and total sunspot area. The radio measurements have an advantage over other indices in that these measurements are completely objective, and can be made under almost any weather conditions. Since they are closely correlated with the magnetic activity which modulates the Sun's energy output with solar irradiance, they correlate closely with other solar activity indices \citep{Covington69}.

The fractal analysis of the solar radio emissions in the frequency range from 245 MHz to 15 GHz has established
that the 2.8 GHz is the most appropriate frequency for the study of the solar coronal rotation \citep{Vats98b, Mehta05}. It is so because the fractal dimension is found to be the least at a frequency near 3 GHz. Therefore, according to \citet{Vats98b} and \citet {Mehta05}, the time series of radio emission data available at 2.8 GHz frequency must have a very strong modulation due to the solar coronal rotation, in comparison to other frequencies. Thus, they used radio emission at 2.8 GHz frequency to determine the coronal rotation period. We, therefore, used this radio frequency to study long term variation in the coronal rotation period, so obtained.

From the knowledge of the yearly mean sunspot numbers (International Sunspot Number, credit: SIDC, 1947-2009) on the photosphere, the years 1954, 1964, 1976, 1986, 1996 and 2009 are the years of solar minima in the solar cycles numbered as 18, 19, 20, 21, 22 and 23, respectively (Figure \ref{srf1}). In fact the year 2006 would have been a solar minima and the number of sunspots would have increased after that, but surprisingly in cycle 23 the minima has become a much extended one. The sunspots number showed a decreasing trend up to the end of the year 2009. It is only now in 2010 that solar activity shows signs of increase. We can see that the radio flux at 2.8 GHz frequency clearly exhibits the 11-years cycle performed generally by the sunspots number on the photosphere. It is well established evidence of a good correlation between the sunspots and the radio emission.

\section{Methods}

There are many statistical methods to analyze such periodic time series. In the present work, the autocorrelation and crosscorrelation method are employed for the analysis of the time series of radio flux at 2.8 GHz.

\begin{figure*}
  \centering{
  \includegraphics[width=154mm]{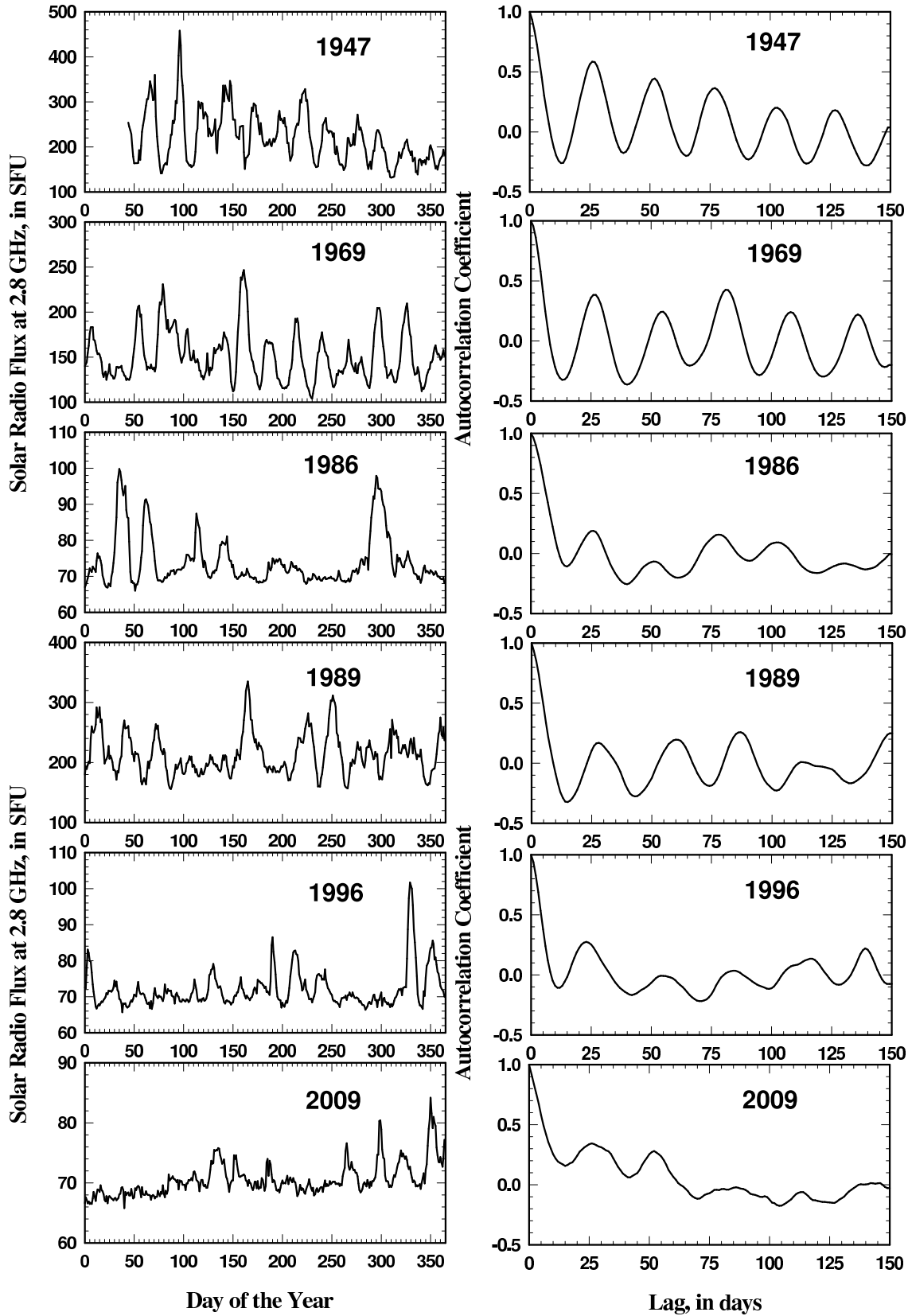}}
 \caption{A few typical time series of solar radio flux at 2.8 GHz (left panel) for the years of solar maxima (1947, 1969 and 1989) and minima (1986, 1996 and 2009) and their respective plots of autocorrelation function used in the present work for the estimation of the synodic rotation period (right panel).}
 \label{srf2}
 \end{figure*}

\subsection{Autocorrelation Analysis}

A time series is defined here as a sequential collection of data observations indexed over time. The series of autocorrelation coefficients represents the correlation between the successive observations of a single time series. The daily measurement of radio flux at 2.8 GHz frequency can be represented in the form of a time series, ($x_0, x_1, x_2, x_3, . . . , x_{n-1}$).

Now we can form $(n-l)$ pairs of observations from this series, where $n$ is the length of the data series (365/366 days in our analysis) and $l$ is referred to as the lag. They are $(x_0, x_l), (x_1, x_{1+l}), (x_2, x_{2+l}), . . . , (x_{n-l-1}, x_{n-1})$.

By regarding the first observation on each pair as one variable and the second observation on each pair as the second variable the autocorrelation coefficient $P_{x}(l)$ as a function of the lag $l$ is given by

\begin{equation}
P_{x}(l) =	P_{x}(-l) = \frac{\sum^{n-l-1}_{k=0} (x_{k} -\overline{x})(x_{k+l} -\overline{x})}{\sum^{n-l}_{k=0} (x_{k} -\overline{x})^2}
\end{equation}

\begin{flushleft}
where $\overline{x}$ is the mean of the daily observations.
\end{flushleft}

The daily solar flux at 2.8 GHz is converted into time series of one year each (total 63 data sets) for the period 1947-2009, each time series beginning with the first day of each calendar year (except for the year 1947, because the data available for this year is from Feb 14 onwards). In this way, each time series of one year length can encompass several cycles of coronal rotation. The data set is almost continuous through out the period of study. We use linear interpolation to fill the few data gaps (less than 1\%) found in some data sets. A few typical examples of such time series are shown in the left panel of the Figure \ref{srf2}. The rotational modulation due to the solar rotation can be noticed in each of the time series. The time series (shown in the left panel of Figure \ref{srf2}) are particularly chosen from the years of solar maxima (1947, 1969 and 1989) and solar minima (1986, 1996 and 2009). The significant rotational modulation of the autocorrelation function in most of the years  is clear display of few nice bright emissive tracers phased well in the longitudes. In some cases more pronounced autocorrelation function can appear even during minima of the solar cycle, if only one isolated (or only very few) active region is on the solar disc for several months.

The autocorrelation function for such time series of the solar radio flux at 2.8 GHz of each year is obtained. The right panels of Figure \ref{srf2} show six such plot of autocorrelation function out of the 63 obtained in the same manner as discussed above. The autocorrelation function shown here corresponds to the years of solar maxima (1947, 1969 and 1989) and solar minima (1986, 1996 and 2009). It can be observed from Figure \ref{srf2} that most of the autocorrelation function curves show a very smooth and cyclic rotational modulation with a fair amount of correlation.

The synodic coronal rotation period can be estimated from the position of the first secondary maximum. To determine the synodic rotation period with maximum possible accuracy, we fitted a sinusoidal wave whose amplitude is the same as first secondary maximum in the autocorrelation function curve and period accurate to a day. Then we considered 7 points (at interval of $\pm 1$ day) around the first secondary maximum of both the sine curve and autocorrelation curve and calculate the standard deviation. Next, we changed the period of sine wave in small steps (in the steps of 0.1 day) and tried to minimize the standard deviation. The value of wave period for which the standard deviation is found to be least would be the synodic rotation period. This fitting allowed us to determine the lag with the sub-day precision and to estimate the uncertainty of the lag determination, which is 0.1 day in present case.

The synodic rotation periods are then converted into sidereal rotation periods using the following expression.

\begin{equation}
T_{sidereal}=\frac{365.26 \times T_{synodic}}{365.26+T_{synodic}}
\end{equation}

This equation does an approximate conversion of synodic to sidereal rotation period. For more accurate conversion one may use the approach outlined by \citet{Rosa95}. Autocorrelation analysis can thus be used to advantage for investigating the periodical variations of time series, particularly the long ones. As we know that the indices of the Sun, derived from the observations or analysis of various solar activity phenomena, also form time series of such type. It is often used with the lag to determine the stationarity of a time series. This analysis is useful in predicting trends or cyclical functions present in a time series. This technique was first employed by \citet{Hansen69} to determine the synodic rotation period of the solar electron corona after analyzing the K-coronameter observations during 1964-1967. Later \citet{El-raey72} analyzed sunspot number, solar radio flux and interplanetary field for the period 1967 to 1970 to determine the solar atmospheric rotation. \citet{Parker82} showed that, within the limit of uncertainty, maximum entropy spectral analysis and autocorrelation analysis provide almost similar results. \citet{Fisher84, Sime89} applied the same analysis technique in their study of the rotational characteristics of white-light and Fe XIV corona, respectively. The differential rotation rates of soft X-ray features in the solar corona were determined by \citet{Weber99} using the Lomb-Scargle periodogram. These results were then compared with the results obtained through the autocorrelation method and were found to agree within $1\sigma$. \citet{Vats98a, Vats98b, Vats01} used the autocorrelation technique to determine the rotation period and the differential rotation as a function of height in the solar corona, respectively.

\begin{figure*}
  \centering{
  \includegraphics[width=\textwidth]{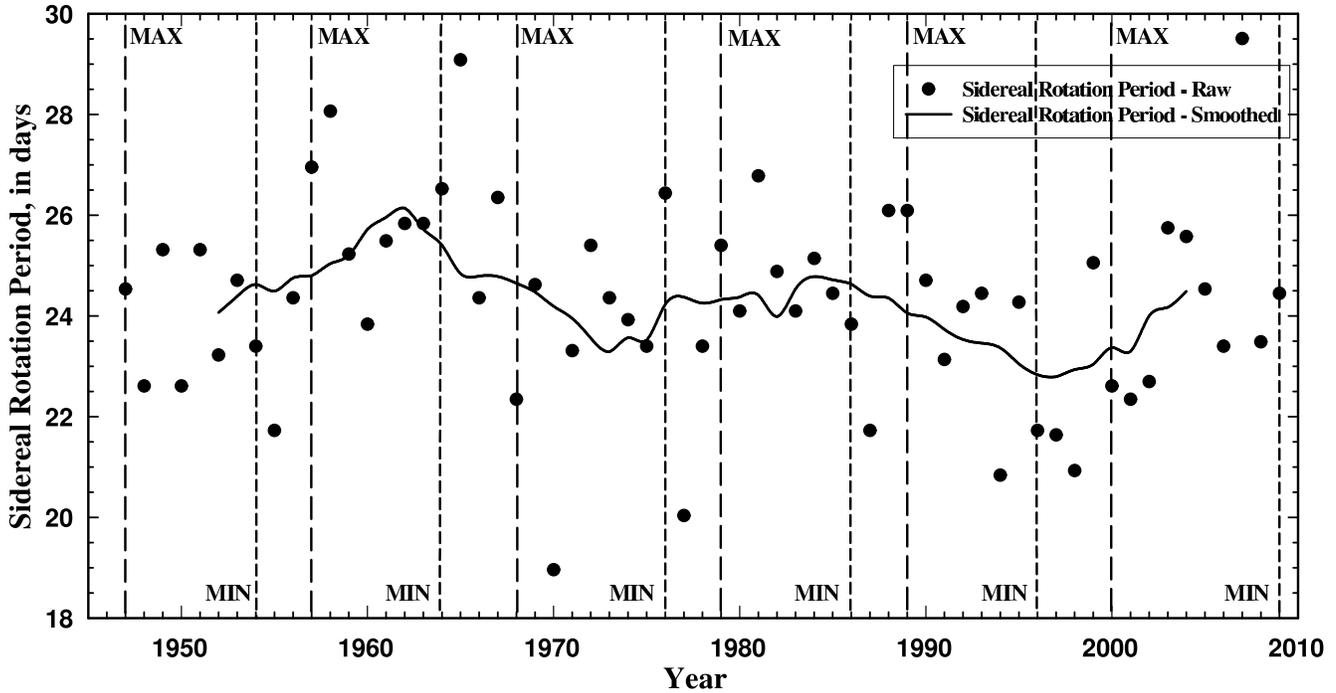}}
 \caption{Temporal variation of the sidereal rotation period obtained using autocorrelation analysis of the solar radio emissions at 2.8 GHz during 1947-2009. The smoothed curve (eleven years running mean) of the sidereal rotation period is also displayed to compare it with its raw results, and also to see, if any long term component  is present in it.}
 \label{srf3}
 \end{figure*}

\subsection{Crosscorrelation Analysis}  

Crosscorrelation can be used to determine the degree of fit between two data sets $x$ and $y$. The day-to-day measurement of annual sunspot number and coronal rotational period, obtained from the analysis of the 2.8 GHz radio flux of the same time interval, can be represented in the form of two time series, $(x_0, x_1, x_2, x_3, . . . . ., x_{n-1})$ and $(y_0, y_1, y_2, y_3, . . . . ., y_{n-1})$.

Again, we can form $(n - l)$ pairs of observations from these two series. They are: $(x_0, x_l)$, $(x_1, x_{1+l})$, $(x_2, x_{2+l})$, . . . . ., $(x_{n-l-1}, x_{n-1})$ and $(y_0, y_l)$, $(y_1, y_{1+l})$, $(y_2, y_{2+l})$, . . . . ., $(y_{n-l-1}, y_{n-1})$.

The crosscorrelation coefficient $P_{xy}(l)$ of the two time series as a function of the lag $l$ is given by,

\begin{equation}
P_{xy}(l) = \frac{\sum^{n-\left|l\right|-1}_{k=0} (x_{k+\left|l\right|} -\overline{x})(y_{k} -\overline{y})}{\sqrt{{\left[\sum^{n-l}_{k=0} (x_{k} -\overline{x})^2\right]}{\left[\sum^{n-l}_{k=0} (y_{k} -\overline{y})^2\right]}}}
\end{equation}

\begin{equation}
P_{xy}(l) = \frac{\sum^{n-l-1}_{k=0} (x_{k} -\overline{x})(y_{k+l} -\overline{y})}{\sqrt{{\left[\sum^{n-l}_{k=0} (x_{k} -\overline{x})^2\right]}{\left[\sum^{n-l}_{k=0} (y_{k} -\overline{y})^2\right]}}}
\end{equation}

\begin{flushleft}
for $l < 0$ and $l > 0$, respectively. Here, $\overline{x}$ and $\overline{y}$ is the mean of the daily observations of two different time series.
\end{flushleft}

The cross-correlation analysis helps in determining the mutual correlation between solar rotation and other observed indices of the Sun as well as other stars. Spectroheliograms obtained in extreme ultraviolet (EUV) lines and the Lyman continuum are used to determine the rotation rate of the solar chromosphere, transition region, and corona \citep{Dupree72, Henze73}. A cross-correlation analysis of the observations indicates the presence of differential rotation through the chromosphere and the transition region. \citet{Bocchialini95} used this method for the diagnostics of the chromospheric dynamics. The correct algorithm to calculate the cross-correlation functions of the He I and the O V line intensities was proposed by \citet{Gomory04}.

\begin{table*}
 \centering
 \begin{minipage}{130mm}
 \caption{Sidereal rotation period (in days) obtained through autocorrelation analysis of daily radio flux at 2.8 GHz for the years 1947-2009.}
  \begin{tabular}{@{}llllllllll@{}}
  \hline
 Year  &Sidereal&Year	 &Sidereal&Year	 &Sidereal&Year	 &Sidereal&Year	 &Sidereal\\
   		 &Rotation&			 &Rotation&			 &Rotation&			 &Rotation&			 &Rotation\\					
   		 &	Period&			 &	Period&			 &	Period&			 &	Period&			 &	Period\\       
  \hline
	1947	&	24.5	&	1960	&	23.8	&	1973	&	24.4	&	1986	&	23.8	&	1999	& 25.1	\\
	1948	&	22.6	&	1961	&	25.5	&	1974	&	23.9	& 1987	& 21.7	& 2000	& 22.6	\\
	1949	&	25.3	&	1962	&	25.8	&	1975	& 23.4	& 1988	& 26.1	& 2001	& 22.3	\\
  1950	&	22.6	&	1963	&	25.8	&	1976	&	26.4	& 1989	& 26.1	& 2002	& 22.7	\\
  1951	&	25.3	&	1964	&	26.5	&	1977	&	20.0	& 1990	& 24.7	& 2003	& 25.7	\\
  1952	&	23.2	&	1965	&	29.1	&	1978	&	23.4	& 1991	& 23.1	& 2004	& 25.6	\\
  1953	&	24.7	&	1966	&	24.4	&	1979	& 25.4	& 1992	& 24.2	& 2005	& 24.5	\\
  1954	&	23.4	&	1967	&	26.4	&	1980	&	24.1	& 1993	& 24.4	& 2006	& 23.4	\\
  1955	&	21.7	&	1968	&	22.3	&	1981	&	26.8	& 1994	& 20.8	& 2007	& 29.5	\\
  1956	&	24.4	&	1969	&	24.6	&	1982	&	24.9	& 1995	& 24.3	& 2008	& 23.5	\\
  1957	&	27.0	&	1970	&	19.0	&	1983	&	24.1	& 1996	& 21.7	& 2009	& 24.4	\\
  1958	&	28.1	&	1971	&	23.3	&	1984	&	25.1	& 1997	& 21.6	& \textbf{Mean}	& 24.3	\\
  1959	&	25.2	&	1972	&	25.4	&	1985	&	24.4	& 1998	& 20.9	& \textbf{std. dev.}& 02.0\\	
  \hline
\end{tabular}
\end{minipage}
\label{tab1}
\end{table*}

\section{Results}

The sidereal rotation period obtained from the time series of the radio flux of the years 1947 to 2009 are tabulated in Table \ref{tab1}. A plot is also given in the Figure \ref{srf3} (by scattered dots), which shows the sidereal rotation period as a function of the year. It is found that the mean sidereal rotation period during 1947-2009 is equal to 24.3 days. The maximum rotation period is found to be 29.5 days in the year 2007 and the minimum rotation period is 19.0 days in 1970. The precision is same in the determination of the sidereal rotation period throughout the period of study and is 0.1 day. The standard deviation of the sidereal rotation period during 1947 to 2009 is calculated and is equal to 2.0 days.

These values of the coronal rotation period show quite a large variability (from 19.0 to 29.5 days). Such large variability in the rotation period has also been reported by \citet{Howard84} in their study of the collection of Mount Wilson white-light plates from 1921 through 1982. From their Table 2, it can be seen that the sidereal rotation rate varies between 11.89 degree/day (in 1965) to 16.47 degree/day (in 1963) in the equatorial region. It means that the sidereal rotation period obtained with white-light plates also shows a large variability from 30.3 to 21.9 days, which is in good agreement with the coronal rotation period obtained in present work. Ironically, the year of maximum and minimum rotation rate (1963 and 1965, respectively) are quite nearby, which again shows analogy with the present work. In some cases, there is a large variation in rotation period in adjacent years (shown in Figure \ref{srf3}).

\citet{Chandra} obtained the coronal rotation period, as a function of latitude, using Nobeyama Radioheliogram (NoRH) images at 17 GHz for the year 1999 through 2005. The sidereal period obtained for the year 2003 shows quite a large variability from 29.2 to 22.1 days. This is again in agreement with our results obtained analizing radio flux data at 2.8 GHz.

To see the presence of any long term variability in it, the coronal rotation period data is smoothen by taking the 3, 5, 7, 9, . . , 17 years running mean of the sidereal rotation period. The smoothed rotation periods so obtained are then plotted against the year. It is found, after comparing all such plots, that the running mean of eleven years smoothed the raw data most. By taking the running mean of eleven years, the short term variation present in the rotation period data sets are smoothed and thus removed. The data obtained after taking the running mean will now show only the long term variation, \textit{i.e.} with the period of more than 11 years. The plot is shown in Figure \ref{srf3} (by solid line). An evidence of long term periodic variation can be seen in this plot. The overall variation has reduced from $\sim 11$ days to only $\sim 3$ days, as shown in Figure \ref{srf3}. This plot has several peaks and dips. However, there are two major maxima and two major minima which can be easily identified in Figure \ref{srf3}. The separation of consecutive maxima or minima is $\sim 22$ years. This is an evidence of $\sim 22$ years periodicity in the temporal variation of sidereal coronal rotation.

\begin{figure}
  \centering{
  \includegraphics[width=80mm]{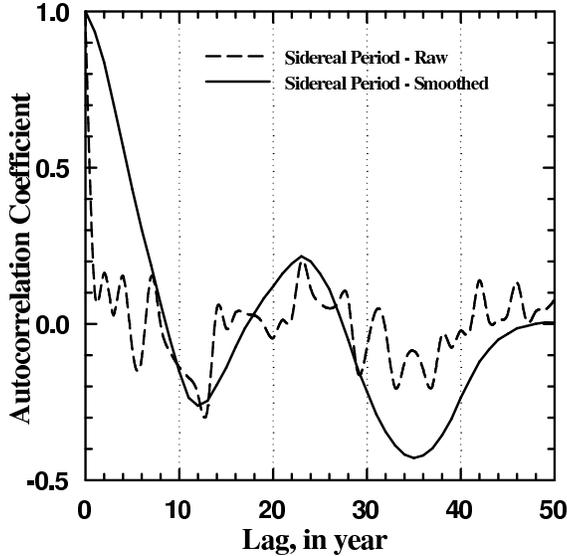}}
 \caption{Autocorrelation function curves of raw and smoothed annual sidereal rotation period for the period 1947-2008.}
 \label{srf4}
 \end{figure}

This periodicity may be better seen in the plot of autocorrelation function. Hence, we derived the autocorrelation coefficient of raw and smoothed (an eleven years running mean) of the sidereal rotation period. These are plotted against the lag in days in Figure \ref{srf4}. The dashed and solid curves are for raw and smoothed sidereal rotation period, respectively. The autocorrelation function exhibits the existence of a cycle close to a period of 22-years. As expected this periodicity is better seen in the autocorrelation of smoothed data set.

\begin{figure}
  \centering{
  \includegraphics[width=80mm]{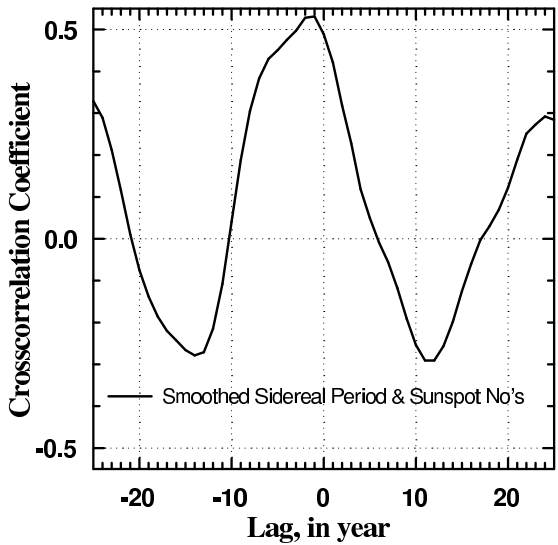}}
 \caption{Crosscorrelation function of smoothed data of annual sidereal rotation period and sunspots numbers.}
 \label{srf5}
 \end{figure}

The crosscorrelation function between the smoothed rotation period and sunspots number is calculated and shown in Figure \ref{srf5} by solid line. It is natural that from a crosscorrelation of these time series, we expect to see evidence of any variation having a period of more than 11 years. The plot of crosscorrelation function in Figure \ref{srf5} (solid line) varies from $\sim -0.30$ to 0.53. One can easily identify two maxima and two minima in this curve and estimate the period. From the positions of these we get a periodicity of $\sim 22$ years. This periodicity in sunspot numbers exists due to the 22-years magnetic cycle. The temporal variation of the coronal rotation period also seems to have a component of the same. However, as the central maximum of the crosscorrelation function fall very close to 0 lag (at -1 year), the smoothed coronal rotation period and sunspot numbers can considered to be almost in phase. 

\section{Discussion}

Based on the model calculations, we know that this radio emission at 2.8 GHz originate from the lower solar corona near $\sim 60,000$ km above the solar surface \citep{Vats01}. Thus the present statistical analysis of solar flux at 2.8 GHz reveals some new long term rotational features of the solar corona at a height $\sim 60,000$ km above the photosphere. The results obtained through the autocorrelation and crosscorrelation analysis can be discussed in the following three parts.

\begin{enumerate}

	\item 
The temporal variation of the rotation period (Figure \ref{srf3}) does not show any systematic periodicity. On the other hand sunspots have a very clear 11 years periodicity. So from this, it seems that probably there is either no or very weak correlation between the coronal rotation period and the sunspot cycle.

	\item
A reasonable presence of a long term variation in the coronal rotation period ( of $\sim$ 22-years) become visible simply by smoothing (11 point running mean) the values (Figure \ref{srf3}). This may correspond to the 22-years magnetic reversal cycle or Hale's cycle.

	\item
The crosscorrelation analysis between the smoothed data of coronal rotation (Figure \ref{srf3}) and its correlation with the sunspots number also shows the clear presence of a 22-years Sun's magnetic field reversal cycle (Figures \ref{srf4} and \ref{srf5}). 

\end{enumerate}

The position of minima in the Figure \ref{srf3} reveals that the years of minima of the 22-years cycle, obtained after smoothing the rotation period data, coincide well with the years of minimum activity of even numbered solar cycles (ENSCs), \textit{i.e.} at the years 1954, 1976 and 1996 (shown in Figure \ref{srf1}). Whereas, the years of maxima of the 22-years cycle matches well with the years of minimum activity (at the years 1964 and 1986) of odd numbered solar cycles (ONSCs). This result is in agreement with the results obtained by \citet{Javaraiah00}. The variation in equatorial rotation rate $A$ is found to be significant only in the years of ONSCs. Where $A$ is reciprocally related to the equatorial sidereal rotation period ($T$) by an expression ($T = 360/A$). It can be seen again in Figure \ref{srf3} (solid line) that the smoothed rotation period rises steeply in the ONSCs and falls back gradually in the ENSCs. A similar trend continues even in solar cycle 23. This shows a good agreement in both the results obtained through two entirely different methods and tracers. 

The Magnetic butterfly diagram due to \citet{Hathaway98} shows clearly that the reversal of the Sun's magnetic polarity is triggered in the years close to the years of solar activity maxima, \textit{i.e.} years 1979, 1989 and 2000. But Figure \ref{srf3} indicates that the ascent or descent in the value of the smoothed sidereal rotation period is initiated in the years of solar minima. Therefore, it shows that there may be anticorrelation between the magnetic pole reversal and the rotation period.

\section*{Acknowledgments}

This work utilizes the radio flux data obtained through the Solar Radio Monitoring Programme, managed by the Dominion Radio Astrophysical Observatory, Penticton, British Columbia, Canada. The international sunspot number is produced by the Solar Influences Data Analysis Center (SIDC) at the Royal Observatory of Belgium. We are indebted to the observers who were involved in the acquisition of the useful solar indices used in the present work. The data (radio flux at 2.8 GHz and sunspot numbers) were acquired from the web page of the National Geophysical Data Center (NGDC). The authors are grateful to Dr. Kiran Biswas for his constructive comments regarding the manuscript. The authors wish to acknowledge the anonymous referee for his valuable suggestions for the improvement of the paper. The research at Physical Research Laboratory (PRL) is supported by the Department of Space, Government of India.


\bsp

\label{lastpage}

\end{document}